\documentclass[12pt,preprint]{aastex}
\catcode`\@=11 
\def\@versim#1#2{\vcenter{\offinterlineskip
        \ialign{$\m@th#1\hfil##\hfil$\crcr#2\crcr\sim\crcr } }}
\def\be{\begin{equation}}
\def\ee{\end{equation}}

\def\lsim{\mathrel{\mathpalette\@versim<}}
\def\gsim{\mathrel{\mathpalette\@versim>}}

\begin{document}

\title{On The Nature of the Variable Infrared Emission from Sgr A*}

\author{Feng Yuan\altaffilmark{1}, Eliot Quataert\altaffilmark{2}, and
Ramesh Narayan\altaffilmark{3}}

\altaffiltext{1}{Department of Physics, Purdue University, West Lafayette,
IN 47907; fyuan@physics.purdue.edu}
\altaffiltext{2}{Astronomy Department, 601 Campbell Hall, University of California, Berkeley, CA 94720; eliot@astron.berkeley.edu}
\altaffiltext{3}{Harvard-Smithsonian Center for Astrophysics, 60 Garden Street, Cambridge, MA 02138; narayan@cfa.harvard.edu}

\begin{abstract}
Recent infrared (IR) observations of the center of our Galaxy indicate
that the supermassive black hole source Sgr A* is strongly variable in
the IR.  The timescale for the variability, $\sim 30$ min, is
comparable to that of the X-ray flares observed by {\em Chandra} and
{\em XMM}, suggesting a common physical origin.  In this paper, we
investigate the nature of the IR emission in the context of models
recently proposed to interpret the X-ray flares. We show that the IR
emission in Sgr A* can be well explained by nonthermal synchrotron
emission if a small fraction of the electrons in the innermost region
of the accretion flow around the black hole are accelerated {into a
broken power-law distribution}, perhaps due to transient events such
as magnetic reconnection. The model predicts differences in the
variability amplitudes of flares in the IR and X-rays, in general
agreement with observations.  It also predicts that the IR emission
should be linearly polarized, as has indeed been observed during one
epoch.  IR and X-ray flares analogous to those observed in Sgr A* may
be detectable from other accreting SMBHs, provided $L \lsim 10^{-8}
L_{EDD}$; at higher luminosities the flaring emission is dominated by
thermal synchrotron-self Compton emission, which is likely to be less
variable.

\end{abstract}


\section{Introduction}

The center of our Galaxy provides the best evidence to date for a
massive black hole (e.g., Sch\"odel et al. 2002; Ghez et al. 2003a),
associated with the compact radio source, Sgr A* (see, e.g., Melia \&
Falcke 2001).  Since the original discovery of Sgr A* in the radio,
there have been intensive searches for counterparts at other
wavelengths, with dramatic progress in the past few years.  Thanks to
the high angular resolution and sensitivity of {\em Chandra} and {\em
XMM-Newton}, Sgr A* has been convincingly detected in the X-rays
(Baganoff et al. 2001, 2003ab; Goldwurm et al. 2003; Porquet et
al. 2003). The X-ray emission has two distinct components.  In
``{quiescence},'' the emission is soft and relatively steady, with a
large fraction of the X-ray flux coming from an extended region with a
diameter $\approx 1.4''$ (Baganoff et al. 2001, 2003b).  Several times
a day, however, Sgr A* has X-ray ``{flares}'' in which the X-ray luminosity
increases by a factor of a few -- 50 for roughly an hour; the short
timescale argues that the emission arises quite close to the BH,
within $\sim 10-100 R_S$ (where $R_S$ is the Schwarzschild radius).
The spectrum of the flares is somewhat uncertain and may be variable;
for the strongest flare detected by {\em Chandra}, the spectrum is
hard, with a photon index of $\Gamma=1.3^{+0.5}_{-0.6}$ (Baganoff et
al. 2001). {\em XMM}, however, detected a very bright and soft flare,
with a photon index of $\Gamma=2.5^{+0.3}_{-0.3}$ (Porquet et
al. 2003).

At IR wavelengths, it has proven extremely difficult to detect Sgr A*
because of contamination by dust and stellar confusion (e.g., Morris
et al. 2001; Genzel et al. 1997; Stolovy et al. 1999; Hornstein et
al. 2002). Recently, however, two groups have independently detected a
highly variable IR source coincident with Sgr A* (Genzel et al. 2003;
Ghez et al. 2003b).  This breakthrough is primarily due to rapid
advances in Adaptive Optics on Keck and the VLT.  Genzel et al. (2003)
detected Sgr A* at 1.6-3.8 $\mu$m (not simultaneously), with a factor
of $\sim 1-5$ variability on timescales of $\sim 10-100$ min.
Similarly, at 3.8 $\mu$m, Ghez et al. (2003b) found that the flux
changes by a factor of 4 over a week, and a factor of 2 in
just 40 min. They argue that the IR spectrum is extremely red, with a
K-L color greater than 2.1 mag.  The rapid variability of the IR
emission argues that it arises from very close to the BH.  Moreover,
the comparable timescales for the IR and X-ray flares strongly
suggests a common physical origin (though rapid IR variability may be
more common than rapid X-ray variability, a point to which we will
return in \S3).

In a previous paper, we described a model for the observed emission
from Sgr A* (Yuan, Quataert, \& Narayan 2003; hereafter YQN03), based
on the idea that accretion onto the central BH proceeds via a hot
two-temperature radiatively inefficient accretion flow (RIAF; Narayan
et al. 1995, 1998; see Narayan 2002, Quataert 2003 for reviews; and
Melia 1992, Liu \& Melia 2001, and Yuan, Markoff \& Falcke 2002 for
related ideas). In this model, as in that of Markoff et al. (2001),
the observed flares are produced when electrons very close to the BH
flow are accelerated to ultrarelativistic energies and emit
synchrotron or synchrotron self-Compton radiation.\footnote{This model
is preferred to one in which the flares arise from fluctuations in the
density or accretion rate, because the latter overpredict the
variability at radio frequencies (Markoff et al. 2001).}  In this
paper, we investigate whether such a model can explain the recently
detected IR emission from Sgr A*. We focus throughout on the RIAF
model.  An alternative possibility that can also account for the
observations of Sgr A* is a coupled RIAF-jet model (Yuan, Markoff \&
Falcke 2002; Markoff et al. 2001; see also Falcke \& Markoff 2000).
More work is needed to distinguish these two possibilities.

\section{RIAF Models for Sgr A*}

As in YQN03, we model the dynamics of the accretion flow as a
two-temperature RIAF.  In this section we briefly summarize the
ingredients in our models.  The dynamical quantities describing the
accreting plasma, such as the density and temperature, are obtained by
solving a one dimensional global model of the accretion flow (see
e.g., Yuan et al. 2000).  Motivated by theoretical calculations and
numerical simulations, we assume that the accretion rate is a function
of radius, with $\dot{M}=\dot{M}_0 (R/R_{\rm out})^s$ (e.g., Blandford
\& Begelman 1999; Stone et al. 1999; Hawley \& Balbus 2002;
Igumenshchev et al. 2000, 2003).  Here $R_{\rm out}$ is the outer
radius of the flow, i.e., the Bondi radius, $\dot{M}_0$ is the
accretion rate at $R_{\rm out}$ (the Bondi accretion rate, fixed by
{\em Chandra} observations of diffuse gas on $\sim 1''$ scales;
Baganoff et al. 2003b), and we fix $s=0.27$ as in YQNO3.  The
suppression of the accretion rate onto the central black hole by a
factor $\sim (R_S/R_{\rm out})^s$ is in good agreement with the high
level of linear polarization detected at 230 GHz (e.g., Aitken et
al. 2000; Bower et al. 2003), which puts an upper limit on the density
of gas near the BH and thus the accretion rate onto the hole (Quataert
\& Gruzinov 2000; Agol 2000).

Electrons in the RIAF have a temperature $\sim 10^{11}$ K close to the
BH, and produce significant emission in the radio by synchrotron
radiation.  In addition to this thermal population of electrons, we
assume that processes such as turbulent acceleration, reconnection,
and weak shocks accelerate some fraction of the electrons into a
power-law tail (as is to be expected in a collisionless magnetized
plasma).  We characterize the nonthermal population by $p$, the slope
of the electron distribution function [$n(\gamma) \propto \gamma^{-p}$
where $\gamma$ is the Lorentz factor], and a parameter $\eta$, the
ratio of the energy in the power-law electrons to that in the thermal
electrons.  The nonthermal electrons emit synchrotron and SSC
radiation from the radio to gamma-rays, depending on the values of $p$
and $\eta$.  In YQNO3, we favored $\eta \sim 1\%$ and $p \ga 3.5$ for
``quiescent'' non-flaring epochs, so as not to violate what were then
IR limits on the emission from Sgr A*.  The {solid} line in Fig. 1
shows a slight variant of this quiescent model with $p = 3$: the IR
flux is now comparable to the lowest flux levels seen in the recent
observations of Genzel et al. (2003) and Ghez et al. (2003b), while
the nonthermal synchrotron emission in the X-rays (dashed line) has a
flux somewhat below that of thermal bremsstrahlung from large radii in
the RIAF (the latter produces the extended source observed most of the
time by {\em Chandra}; e.g., Quataert 2002).

\section{Variable IR and X-ray Emission from Accelerated Electrons}

The emission produced by the nonthermal electrons depends sensitively
on the uncertain electron distribution function, parameterized by $p$
and $\eta$ (defined earlier), as well as the maximum Lorentz factor of
the nonthermal electrons, $\gamma_{\rm max}$.  In addition, the relative
importance of different acceleration mechanisms may be time variable,
which would result in highly time variable nonthermal emission.  A
useful analogue may be solar flares, in which the amount of energy in
suprathermal electrons differs by several orders of magnitude from
flare to flare (see, e.g., Fig. 3 of Miller et al. 1997).  In the
following, we investigate how accelerated electrons close to the BH
can account for the recently detected variable IR emission from Sgr
A*, and how such IR emission is related to the X-ray flares.  We also
stress how relatively modest changes in the electron distribution
function can produce a wide range of variability amplitudes and
spectra in the IR and X-ray.

\subsection{Synchrotron self-Compton (SSC) models}

In YQN03, we showed that if a large fraction of the electrons ($\eta
\sim 1$) are accelerated into a power-law tail with $\gamma_{\rm max}
\sim 10^3$, the accelerated electrons will Compton-scatter synchrotron
photons into the {\em Chandra} band, producing an X-ray flare.  We
explored two such synchrotron self-Compton (SSC) models with different
values of the electron energy index: $p=0.5$ and $p=1.1$.  In both
models, we predicted that there should be a significant flare in the
IR correlated with the X-ray flare (see Figs. 7a,b in YQN03).  Fig. 2
of the present paper shows our $p = 1.1$ model as compared with the
recent IR detections.  Because of the extreme variability that is
observed, and the lack of simultaneous X-ray and IR detections, we do
not attempt to distinguish rigorously between a ``quiescent'' and
``flaring'' IR source at this point, though the smallest and largest
IR fluxes detected should be a useful guide.  In addition, because
there are no simultaneous IR observations in several bands, we do not
compare our predicted IR spectra with the observations (see Ghez et
al. 2003 and Genzel et al. 2003 for possible constraints on spectra
with non-simultaneous observations).

As Fig. 2 shows, the predicted IR emission in the SSC model is in good
agreement with observations of the Galactic Center.  The difficulty
with this model, however, is the rather extreme requirement that $\eta
\sim 1$, i.e., nearly all of the electron energy should reside in a
very hard power-law tail.  There is also no well-motivated reason for
$\gamma_{\rm max} \lsim 10^3$, as is required to avoid significantly
overproducing the IR flux (since acceleration mechanisms such as
reconnection, turbulence, or shocks could readily accelerate electrons
to much higher energies).  For these reasons, it is worth exploring
models which relax the above assumptions on $\eta$ and $\gamma_{\rm
max}$.  The nonthermal X-ray emission must then be due to synchrotron
radiation, not SSC, and so we refer to such models as ``synchrotron''
models.

\subsection{Synchrotron models}

In the synchrotron model for X-ray flares described in YQN03, we
assumed that in the inner $\approx 10 R_S$ of the accretion flow a
small fraction of electrons are accelerated into a hard power-law tail
with $\gamma_{\rm max} \sim 10^6$ or larger. The synchrotron emission
from such high energy electrons accounts for the observed X-ray
flares.  The specific models considered in YQN03 modeled the
distribution of accelerated electrons as a single power-law with $p
\sim 1$.  In this case the flare contains too few electrons with
$\gamma \sim 10^3$ to produce significant IR flux via synchrotron
emission; therefore, the model does not produce an IR flare (see,
e.g., Fig. 6 of YQN03).  The dashed lines in Fig. 3 show, however,
that a model in which the accelerated electrons have $\eta \approx
7\%$ and $p \approx 2$ can produce an IR flux consistent with the
observations, while at the same time producing an X-ray flux similar
to that observed during flares by {\em Chandra} and {\em XMM}.  This
model has the advantage of theoretical economy: the same population of
power-law electrons produces both the IR and X-ray flares.  One
problem, however, is that the predicted X-ray spectrum ($\Gamma
\approx 2$) is too soft to be consistent with the X-ray flares
observed by Chandra.

Thus, if we wish to fit the IR flux, the X-ray flux, and the spectra
of flares with a synchrotron model, it appears that a single power-law
distribution of accelerated electrons is ruled out and a two-component
distribution is favored. In fact, such a two-component distribution is
expected {\it theoretically} from the physics of particle
acceleration, either in shocks or via magnetic reconnection. For
example, models of nonlinear shock acceleration (which account for the
back reaction of the accelerated particles on the shock structure)
typically predict that the spectral index of accelerated particles
increases (``hardens'') with increasing energy (e.g., Ellison et
al. 2000); as discussed below, this kind of hardening is what we need
to account for the observations of Sgr A*.  More generally, although
the physics of particle acceleration in the accretion flow context is
not that well understood, it is nevertheless quite likely that, while
some particles are {\em accelerated} into a hard power-law
distribution, many more particles are {\em heated} to a roughly
thermal distribution (see the discussion and references in \S5 of
YQN03).  There is observational support for this result both for shock
acceleration (e.g., from the study of SNR shocks; see McKee \&
Hollenbach 1980; Vink 2003) and magnetic reconnection (from the study
of solar flares; see Priest \& Forbes 2000; Lin \& Johns 1993).  Many
details remain unclear but current results seem to indicate that the
energy distribution of the power-law electrons is harder in the case
of magnetic reconnection compared to shock acceleration.  Since we
need a very hard distribution ($p_2 \sim 1$, see below), we feel that
reconnection is a more likely explanation for the flares in Sgr A*.

{For convenience, we model the lower energy (``heated'') electrons in
the flare by means of a steep power-law ($p_1\sim3$).  We have also
calculated models in which the heated electrons are described by a
thermal distribution with a higher temperature than the bulk of the
electrons, and we obtain similar results.  This is because the precise
form of the energy distribution for the low-energy electrons is not
that well constrained without detailed IR spectra; current
observations require only that an appropriate number of electrons with
IR-emitting Lorentz factors should be present.  In contrast, the
distribution of the high-energy power-law electrons is more important
since it directly determines the spectral index of the flare radiation
in the X-ray band.


Motivated by these considerations, we assume that, episodically, some
fraction of electrons at $R \la 10R_S$ are accelerated into the
following ``broken'' power-law distribution,
\begin{mathletters}
\be n_{\rm pl}(\gamma)=N_{\rm pl}(p_1-1)\gamma^{-p_1}, \hspace{1cm}
\gamma_{\rm min} \le \gamma \le \gamma_{\rm mid}, \ee \be n_{\rm
pl}(\gamma)=N_{\rm pl}(p_1-1)\gamma_{\rm mid}^{p_2-p_1}\gamma^{-p_2},
\hspace{1cm} \gamma_{\rm mid} \le \gamma \le \gamma_{\rm max}. \ee
\end{mathletters}
The above prescription refers to the injected distribution. For the
steady distribution of electrons, the power-law will become steeper
above a cooling break at $\gamma_c$. We calculate the values of
$\gamma_c$ and $\gamma_{\rm min}$ self-consistently at each radius as
described in YQN03. For the reasons explained above, we adopt $p_1=3$
and $p_2=1$; the former corresponds roughly to the inferred slope from
the quiescent emission while the latter corresponds to that inferred
from the hard X-ray flares.  To determine the break between these two
distributions, which occurs at $\gamma_{\rm mid}$, we define
$\eta_{IX}$ to be the ratio of the energy in electrons with $p = p_1$
to those with $p = p_2$.  Since the electrons with relatively small
(large) $\gamma$ are responsible for the IR (X-ray) radiation,
$\eta_{IX}$ is essentially the ratio of the energy in IR and X-ray
emitting electrons.

The solid lines in Fig. 3 show the spectrum from such a population of
electrons for $\eta = 7\%$, $\gamma_{\rm max} \sim 10^6$, and
$\eta_{IX} = 1$ (comparable energy in IR and X-ray emitting
electrons). {This corresponds to $\gamma_{\rm min} \sim 100-500$ and
$\gamma_{\rm mid} \sim 10^5$ in the region $R \la 10R_S$.  Although
the power-law electrons have $\approx 7 \%$ of the electron thermal
energy, the total fraction of the electrons accelerated is only $\sim
0.5\%$, while the ratio of electrons with $p\sim 1$ to those with
$p\sim 3$ is $\sim 2\%$. {The total number density of electrons in
this region is $\sim 10^7$ cm$^{-3}$ and the strength of the magnetic
field is $\sim 20-100$G}. The thin solid line in Fig. 3 shows the
flaring synchrotron emission from the ``broken'' power-law electrons
while the thick solid line shows the sum of the flaring emission and
the quiescent radiation.  We see that the model produces a large IR
flux comparable to that observed, as well as a hard X-ray flare
similar to that seen by {\em Chandra}.


  
Because the physics of electron acceleration in the collisionless
magnetized plasma close to the BH remains poorly understood it is
interesting to assess how the predicted spectrum changes when some of
our assumptions are varied.  This is shown in Fig. 4.  The thin and
thick dashed lines show the results of changing the relative energy in
IR and X-ray emitting electrons: $\eta_{IX}=0.1$ (thick dashed, X-ray
dominated) and $\eta_{IX}=10$ (thin dashed, IR dominated).  In the
$\eta_{IX} = 0.1$ case, there is a prominent X-ray flare with little
change in the IR emission, while for $\eta_{IX} = 10$, the X-ray flare
is much less prominent (and would in fact disappear entirely for
$\eta_{IX} \gg 10$).  In general, X-ray flares are predicted to have
much larger amplitudes than IR flares, in part because the baseline
``quiescent'' flux is substantially smaller in X-rays.  This trend is
consistent with current observations.

Fig. 4 also explores the effects of changing $\gamma_{\rm max}$.  The
thick solid line shows the spectrum when $\gamma_{\rm max}$ is 5 times
smaller than in Fig. 3, namely $\gamma_{\rm max} \sim 2 \times 10^5$.
In this case the peak of the synchrotron emission is in the X-ray
band. Such a spectrum could account for the bright {\em soft} X-ray
flare detected by {\em XMM} (Porquet et al. 2003).  The thin solid
line shows the result of further decreasing $\gamma_{\rm max}$ to
$\sim 3 \times 10^4$.  In this case, there is still a large increase
in IR flux, but there is no X-ray flare because the peak of the
synchrotron emission occurs well below the X-ray band.  This
possibility, namely that it is easier to accelerate low energy
electrons that emit in the IR, could account for the fact that
significant IR variability appears more common than significant X-ray
variability (\S1).

\section{Summary and Discussion}

Recent observations have convincingly detected IR emission from Sgr
A*, the supermassive black hole in the Galactic Center.  The emission
is highly variable, with order unity changes on timescales $\sim 30$
min, comparable to the orbital period of matter near the last stable
orbit around the BH.  The amplitude of the IR variability is $\la 5$,
smaller than that of the X-ray flares observed by {\em Chandra} and
{\em XMM}. The timescales are, however, quite comparable, strongly
suggesting that the IR and X-ray emission arise from the same physical
location.  In YQN03 we presented flare models based on the SSC
mechanism that predict roughly the right flux in both the IR and X-ray
bands (see Fig. 2).  In this paper, we have explored in more detail an
alternative model in which synchrotron emission from electrons
accelerated in the accretion flow close to the BH account for the
variable IR and X-ray emission (Fig. 3). {In order to fit both the IR
and X-ray observations, we assume that the accelerated electrons are
in a broken power-law distribution, with a hardening of the power law
at high energies.  We discuss possible physical origins for such a
distribution in \S3.2}. In all of our models, the IR emission is
produced by the synchrotron process and is predicted to be
significantly polarized.  This finds support in the high level of
linear polarization detected in one observation by Genzel et al. (The
IR flux was relatively low during this epoch so they identified it as
``quiescence'').  It is worth noting that Rees et al. (1982), in their
original ion-torus model, proposed that the prominence of nonthermal
emission would be a signature of two-temperature collisionless
accretion flows; we suggest that this prediction has been borne out by
the recent observations of the Galactic Center.

We have also explored in Fig. 4 the range of IR and X-ray spectra
expected as a result of flare-to-flare variations in the distribution
function of accelerated electrons (as is observed in, e.g., solar
flares).  We suggest that such variations can account for the wide
range, both in flux and spectra, of IR and X-ray emission from Sgr A*.
Our calculated models make several predictions: (i) X-ray and IR
flares should often be correlated, but need not always be (see
Fig. 4).  (ii) X-ray flares, when present, should have larger
amplitudes than the corresponding IR flares.  (iii) The spectral
slopes of X-ray flares can differ significantly, whereas the slopes of
IR flares should, at least within our simple models, differ less.
(iv) X-ray and IR flares should be accompanied by only small amplitude
variability in the radio and sub-mm (because of the damping effect of
the thermal electrons; see \"Ozel, Psaltis \& Narayan 2000). 
In fact, these predictions are probably generic to many
nonthermal models (e.g., even if the acceleration is in a jet rather than
the accretion flow). In this
connection, a distinction should be made between the short time scale
flares, which we identify with localized acceleration of nonthermal
electrons, and longer time scale variability which may be caused by
variations in the temperature, density and magnetic field strength
close to the BH.  The latter variations will cause modest amplitude
variability in all bands, including radio/submm, but on longer time
scales.

The remarkable detection of short-timescale variability in the IR and
X-ray emission from the Galactic Center raises the important question
of why such flares have not been detected from other systems, given
the plausible hypothesis that most low-luminosity AGN accrete via a
RIAF (Narayan 2002; Quataert 2003). One factor is certainly the large
amount of observing time dedicated to Sgr A*, but this may not be the
only explanation.  To assess this issue, Fig. 5 shows the quiescent
(thick lines) and flaring (thin lines) spectra for accretion rates 10,
100, and 1000 times that of Sgr A*.  In each calculation, we assume
that the basic accretion flow parameters and electron distribution
function ($s$, $\eta$, $\eta_{IX}$, etc.)  are the same as they are in
our synchrotron model of Sgr A* (Fig. 3), except that the accretion
rate onto the BH is larger.  This assumes that the basic acceleration
physics is independent of $\dot M$ (e.g., density), which may not be
correct but is a useful ``straw-man'' model.  The results of Fig. 5
show that for more luminous systems the flaring emission becomes much
more difficult to observe.  If we scale to Eddington units ($L_{\rm Edd}
\sim 4\times10^{44} ~{\rm erg\,s^{-1}}$ for Sgr A*), the flares become
difficult to detect for $L \gsim 10^{-8} L_{\rm Edd}$ because they are
swamped by the larger quiescent emission.  The physical reason is that
thermal SSC emission becomes much more important than nonthermal
emission as the accretion rate increases.  This emission would
undoubtedly be variable because of fluctuations in, e.g., $\dot M$,
but the resulting variations in flux would likely only be of order
unity.  As a result, we expect that dramatic flaring analogous to that
seen in Sgr A* will be restricted to very low-luminosity systems ($L
\lsim 10^{-8} L_{Edd}$) where the nonthermal emission can dominate the
thermal SSC emission. 

To be more concrete, we point out several systems where the results of
Fig. 4 suggest that flaring may be detectable. This discussion is not
exhaustive, but highlights some of the relevant issues in searching
for variability analogous to that seen in Sgr A*.  One possibility is
M84, which has two reported nuclear X-ray luminosities: $L_X \approx
10^{39}$ ergs s$^{-1}$ (Ho et al. 2001) and $L_X \approx 4 \times
10^{39}$ ergs s$^{-1}$ (Finoguenov \& Jones 2001).\footnote{M84 has an
X-ray emitting jet so it is not clear whether this emission arises
from the direct vicinity of the black hole.} These correspond to
$\approx 1-4 \times 10^{-8} L_{Edd}$ for the $M \approx 1.5 \times
10^9 M_\odot$ black hole in this system.  One complication is that the
relevant dynamical times close to a BH scale linearly with black hole
mass.  Hour-long flares $\sim$ once per day in Sgr A* thus correspond
to $\sim 1$ flare per year for $M = 1.5 \times 10^9 M_\odot$, with a
duration of 1-2 weeks.  It is unlikely (few \% probability) that a
random observation would see such flares.  A sample of $\sim 25$
comparable mass systems observed several times over the course of a
year could, however, detect $\sim 1$ flare.  A more promising
candidate is M32, where Ho et al. (2003) found a $10^{36}$ ergs
s$^{-1}$ X-ray source coincident with the $2.5 \times 10^6 M_\odot$
black hole.  This X-ray luminosity corresponds to $\approx 3 \times
10^{-9} L_{Edd}$, implying that M32 may show X-ray flares according to
our analysis.  The timescales are expected to be similar to those in
Sgr A* since the black hole masses are similar.  Finally, if we
consider the other extreme of low-luminosity X-ray binaries, a
possible candidate for the detection of X-ray flares would be XTE
J1118+480, or its twin A0620--00. {\em Chandra} observations show that
$L_X\sim 10^{-8.5}L_{\rm Edd}$ (McClintock et al. 2003).  The flare
duration in this source would, however, be $\sim$ 0.01 second, and the
flux only $\sim 4 \times 10^{-13} \ {\rm ergs\,cm^{-2}\,s^{-1}}$.
Unfortunately, it is not currently possible to detect such short
duration flares from these very faint X-ray binaries.  We thus
conclude that very low-luminosity AGN with black hole masses
comparable to or even somewhat smaller than that of Sgr A* are the
most promising sources to search for rapid large-amplitude variability
analogous to that seen in the Galactic Center.



\acknowledgements It is a pleasure to thank Reinhard Genzel for
conversations on IR observations of Sgr A*. We also thank our referee,
Heino Falcke, for his helpful suggestions and comments. This work was
supported in part by NSF grants AST 0307433, AST 0206006, NASA grants
NAG5-9998, NAG5-10780, NAG5-12043, and an Alfred P. Sloan Fellowship
to EQ.

\newpage

\begin{figure}
\epsscale{0.90}
\plotone{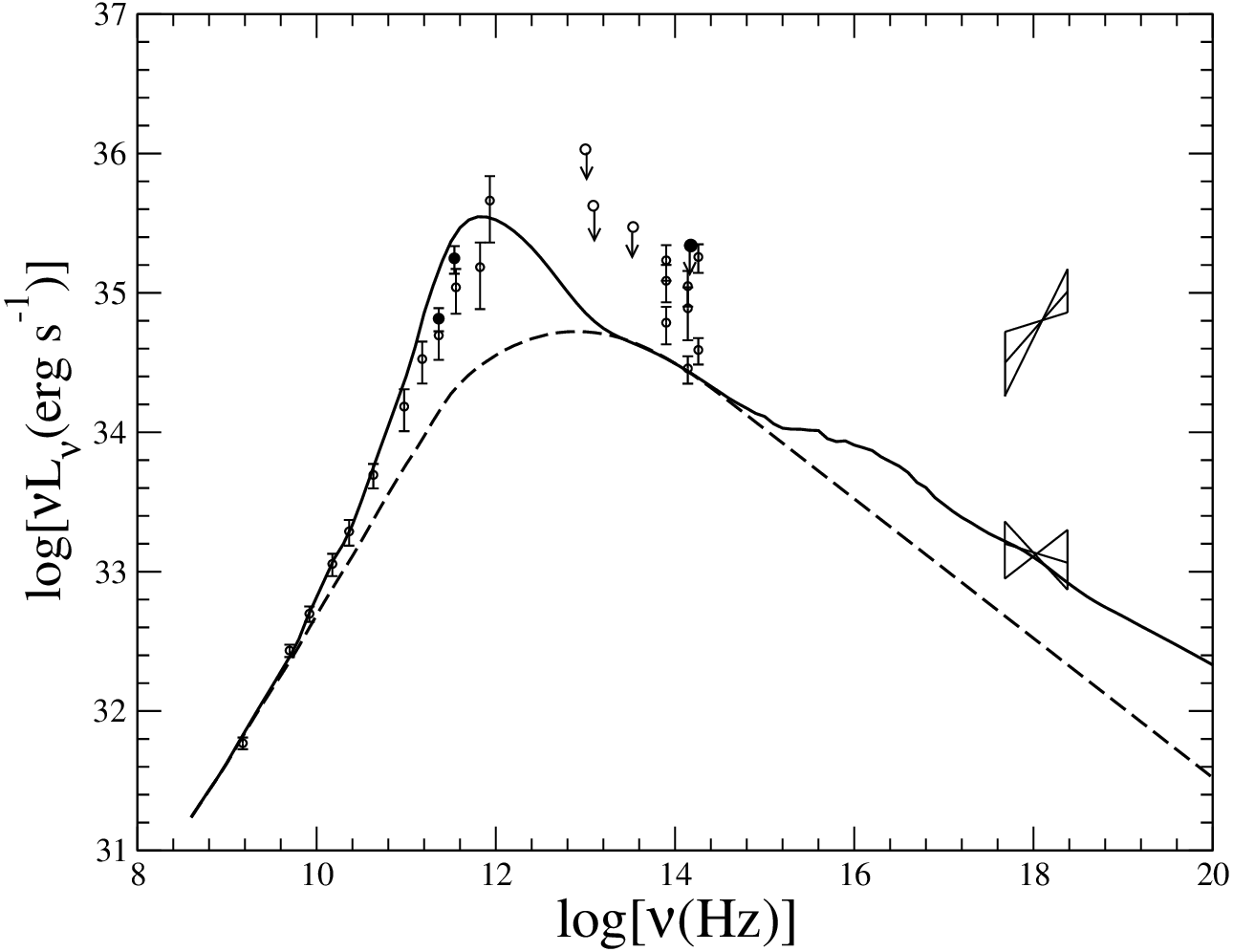}
\vspace{.3in}
\caption{RIAF model for the quiescent state of Sgr A*.  The IR data
with error bars are from Ghez et al. (2003b) and Genzel et al. (2003),
the radio data from Falcke et al. (1998, open circles) and Zhao et
al. (2003, filled circles), the far-IR data with upper limits from
Serabyn et al.  (1997) and Hornstein et al. (2002), and the two
``bowties'' in the X-ray for the quiescent (lower) and flaring
(higher) states from Baganoff et al. (2003b, 2001).  The dashed line
shows the synchrotron emission by power-law electrons with $p=3$.  The
solid line shows the total quiescent emission, including that from
thermal electrons. The slight difference in the value of $p$ compared
with that in YQN03 ($p=3.5$) is to fit the quiescent IR data better.}
\end{figure}

\begin{figure}
\epsscale{0.90}
\plotone{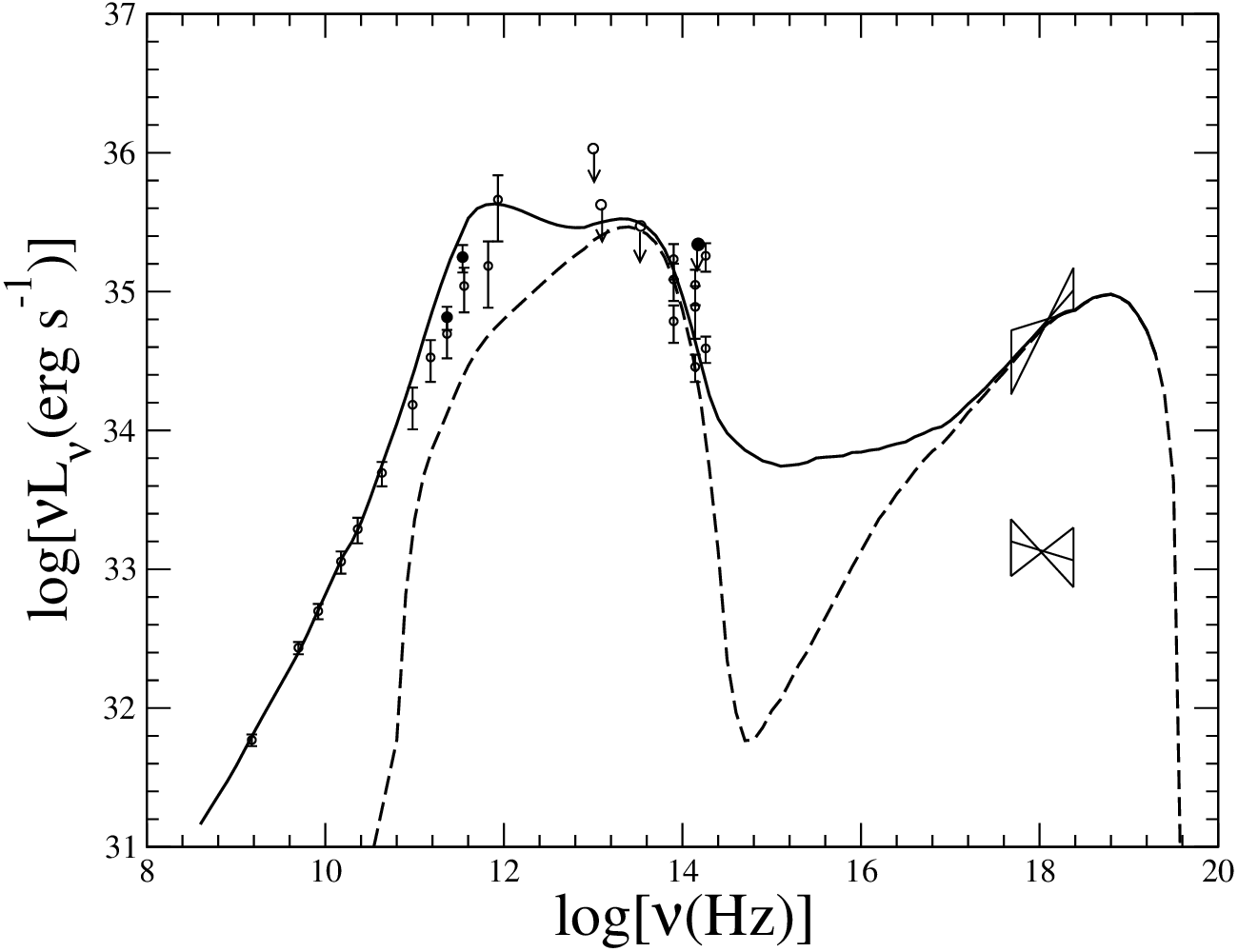}
\vspace{.3in}
\caption{Synchrotron-self-Compton (SSC) model for the IR and X-ray
flares in Sgr A* (from YQN03): The dashed line shows the synchrotron
and SSC emission from $p=1.1$ power-law electrons, and the solid line
shows the total flare spectrum, including the emission from the
thermal electrons.}
\end{figure}

\begin{figure}
\epsscale{0.90}
\plotone{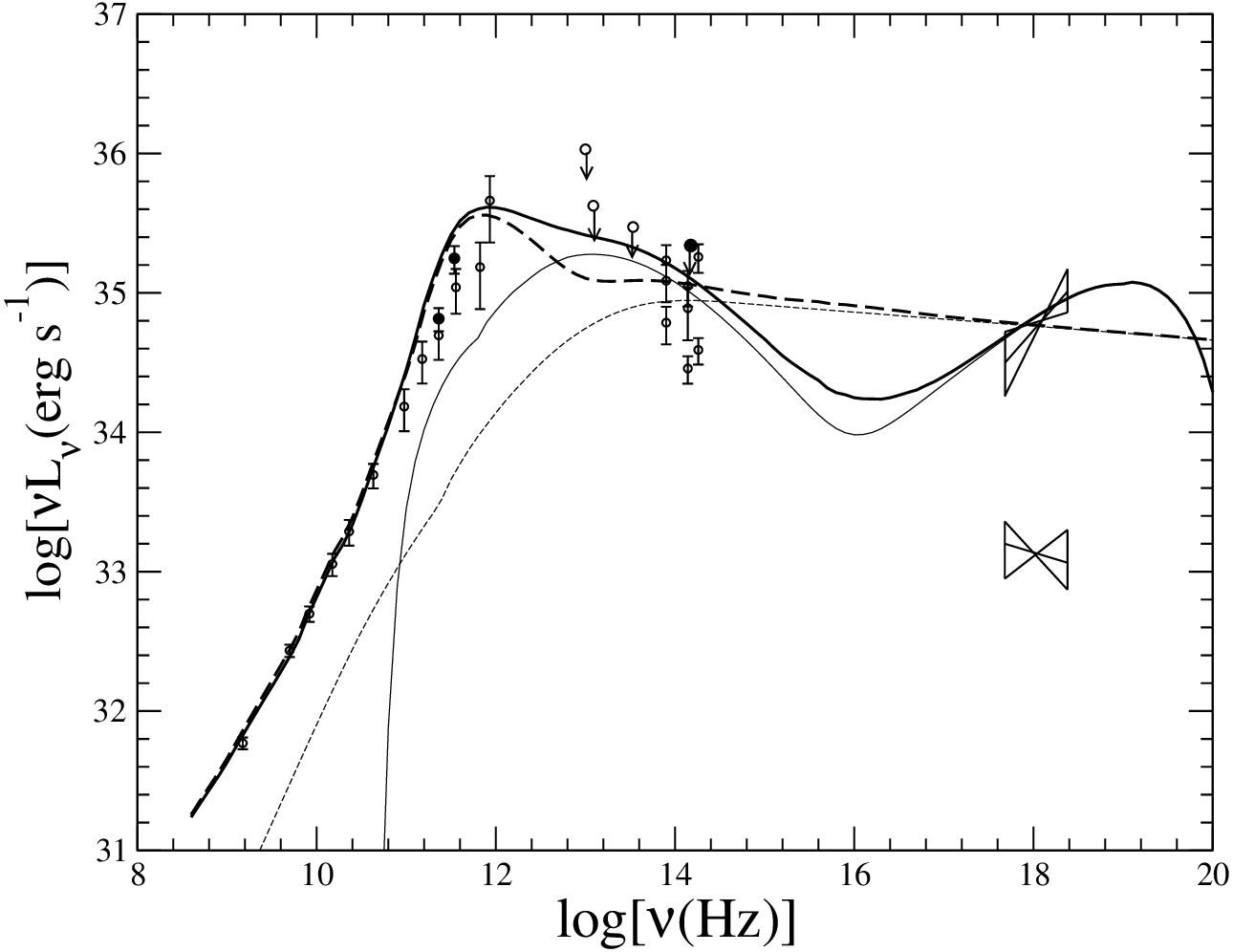}
\vspace{.3in}
\caption{Pure synchrotron models for the IR and X-ray flares in Sgr
A*.  The two dashed lines are models in which the electrons are
assumed to have $p=2.1$.  The solid lines are for the ``broken
power-law'' model (eq. 1), with $p_1=3$, $p_2=1$, $\eta=7\%$,
$\gamma_{\rm max} \sim 10^6$ and $\eta_{IX}=1$.  In each case, the
light lines correspond to the emission from only the power-law
electrons and the thick lines to the total emission including the
thermal electrons.}
\end{figure}

\begin{figure}
\epsscale{0.90}
\plotone{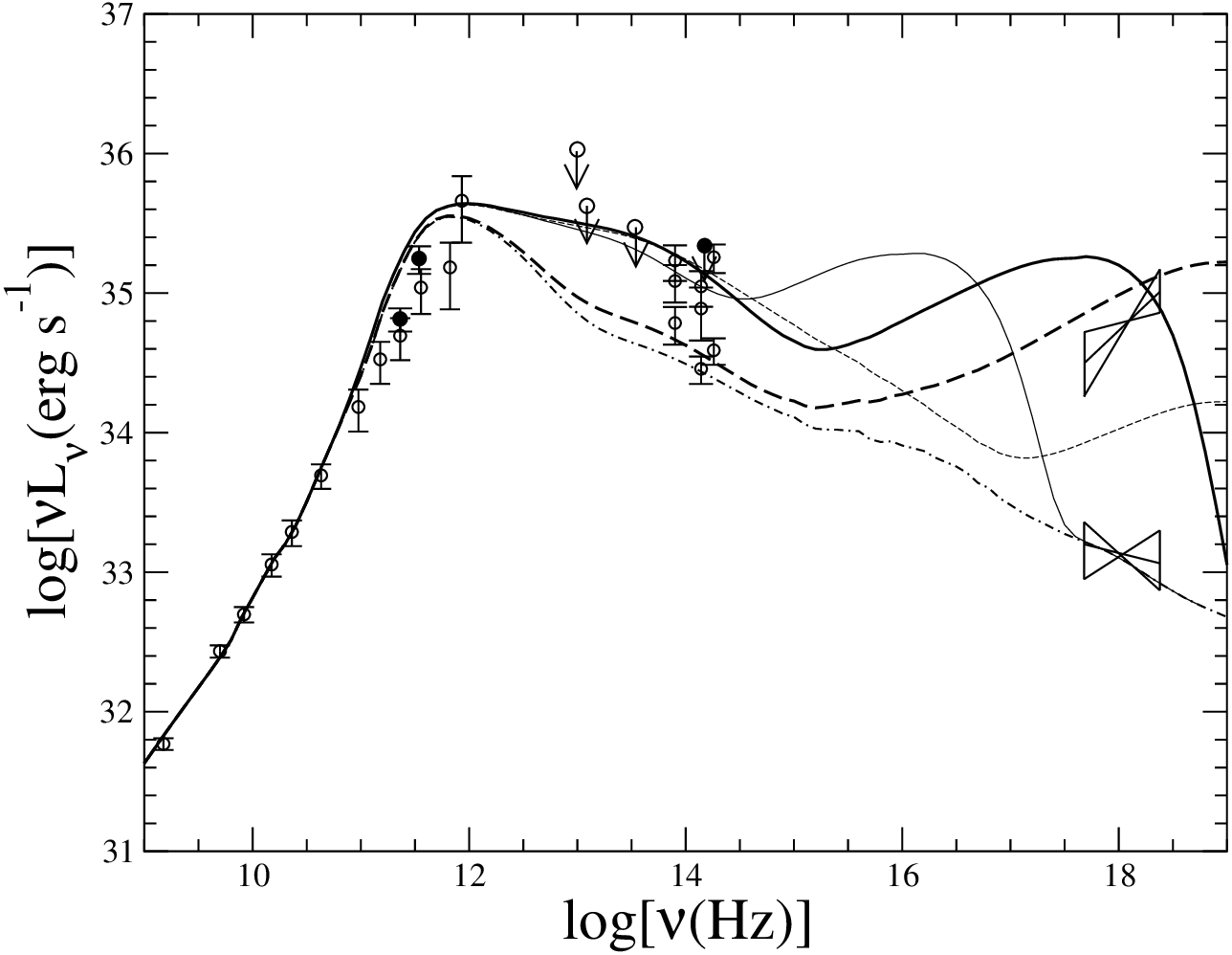}
\vspace{0.2in}
\caption{Effect of varying the parameters of the broken power-law
synchrotron model (thick solid line) shown in Fig. 3.  The two dashed
lines show the effect of varying $\eta_{IX}$, keeping $\eta=5\%$ and
$\gamma_{\rm max} \sim 10^6$ fixed.  The thin dashed line corresponds
to $\eta_{IX}=10$ (IR dominated) and the thick dashed line to
$\eta_{IX}=0.1$ (X-ray dominated).  The two solid lines show the
effect of varying $\gamma_{\rm max}$, keeping $\eta=10\%$ and
$\eta_{IX}=1.0$ fixed.  The thick solid line corresponds to
$\gamma_{\rm max} \sim 2\times10^5$ and the thin solid line to
$\gamma_{\rm max} \sim 3\times10^4$.  The dot-dashed line is the model
of the quiescent state shown in Fig. 1.}
\end{figure}

\begin{figure}
\epsscale{0.90}
\plotone{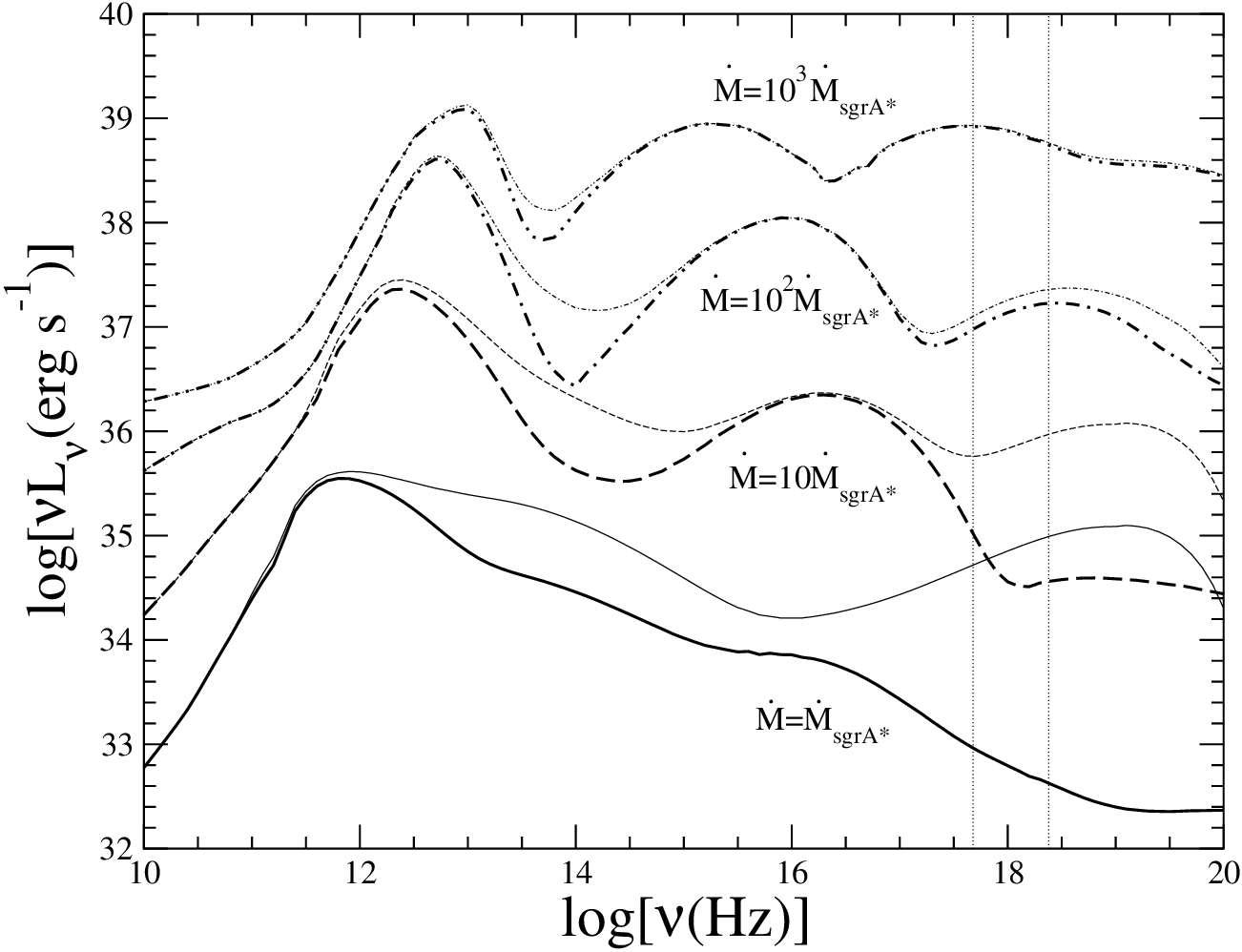}
\vspace{0.2in}
\caption{Quiescent (thick lines) and flaring (thin lines) spectra for
different accretion rates. The solid lines correspond to the broken
power-law synchrotron model of Sgr A* shown in Fig. 3.  The dashed,
dot-dashed, and dot-dot-dashed lines are for systems with accretion
rates of 10, 100, and 1000 times the rate in Sgr A*.  For more
luminous systems (higher $\dot M$), the SSC emission from thermal
electrons increases substantially.  As a result, the emission from
flares would be much more difficult to detect.}
\end{figure}

\end{document}